\documentclass[12pt]{article}
\usepackage{cite}
\usepackage{amssymb}
\usepackage{amsmath}
\numberwithin{equation}{section}

\def\Vol{\mathrm{Vol}}
\def\dist{\mathrm{dist}}

\def\s{\sigma}

\def\nn{\nonumber}

\def\BR{{\mathbb R}}

\newtheorem{Pa}{Paper}[section]
\newtheorem{Tm}[Pa]{{\bf Theorem}}

\newtheorem{Cy}[Pa]{{\bf Corollary}}
\newtheorem{Rk}[Pa]{{\bf Remark}}

\newtheorem{Ee}[Pa]{{\bf Example}}
\newtheorem{Dn}[Pa]{{\bf Definition}}
\newtheorem{Pn}[Pa]{{\bf Proposition}}
\title{The Boltzmann equation and corresponding extremal problems}
\author{Lev Sakhnovich}
\date{}
\begin{document}
\maketitle

\thanks{99 Cove ave., Milford, CT, 06461, USA \\
 E-mail: lsakhnovich@gmail.com}\\
 
 
\begin{abstract}
We start with some global
Maxwellian function $M$, which is a stationary solution (with the constant total density $\rho$) of the Boltzmann equation, and we denote the number of the corresponding space variables by $n$.
The notion of distance between the global
Maxwellian function  and an arbitrary solution $f$
(with the same  total density $\rho$ at the fixed moment $t$) of the Boltzmann equation is introduced.
In this way we essentially generalize the important Kullback-Leibler distance, which was used
before. An extremal problem to find a solution of the Boltzmann equation, such that $\dist\{M,f\}$ is minimal in the class of solutions with the fixed values
of energy and of $n$ moments, is solved.
\end{abstract}

\section{Introduction.} 
 The well-known Boltzmann equation for the monoatomic gas has the form
 \begin{equation}\label{1}
 \frac{\partial{f}}{\partial{t}}=-\zeta{\cdot}\triangledown_{x}f+
 Q(f,f),\end{equation}
 where  $t{\in}\BR$ stands for time,   $x=(x_{1},...,x_{n}){\in}\Omega$ stands for
 space coordinates,  $\zeta=(\zeta_{1},...,\zeta_{n}){\in}\BR^{n}$ is velocity,
 and $\BR$ denotes the real axis. The collision
 operator $Q$ is defined by the relation
 \begin{equation}\label{1'}
 Q(f,f)=\int_{R^{n}}\int_{S^{n-1}}[f(\zeta^{\prime})f(\zeta^{\prime}_{\star})-
 f(\zeta)f(\zeta_{\star})]B(\zeta-\zeta_{\star},\sigma){d\sigma}d\zeta_{\star},
 \end{equation}
 where $B(\zeta-\zeta_{\star},\sigma){\geq}0$  is the collision kernel.
 Here we used the notation
 \begin{equation}\label{2}
 \zeta^{\prime}=(\zeta_{\star}+\zeta)/2+\sigma|\zeta_{\star}-\zeta|/2,\,
 \zeta^{\prime}_{\star}=(\zeta_{\star}+\zeta)/2-\sigma|\zeta_{\star}-\zeta|/2,
 \end{equation}where $\sigma{\in}S^{n-1}$, that is, $\sigma{\in}\BR^{n}$
 and $|\s|=1$.
 The solution $f(t,x,\zeta)$ of Boltzmann equation \eqref{1} is the distribution function of gas. We start with some global
Maxwellian function $M$, which is the stationary solution (with the total density $\rho$) of the Boltzmann equation.
The notion of distance between the global
Maxwellian function  and an arbitrary solution $f$
(with the same  value  $\rho$ of the total density at the fixed moment $t$) of the Boltzmann equation is introduced.
In this way we essentially generalize the Kullback-Leibler distance 
\cite{KL}, which was fruitfully used
before (see further references in the recent papers \cite{Hab, SH, Vi2}).
Our approach enables us to treat also the non-homogeneous case.
 An extremal problem to find a solution of the Boltzmann equation, such that 
 $\dist\{M,f\}$ is minimal in the class of solutions with the fixed values
of energy and of $n$ moments, is solved.

Some  necessary preliminary definitions and results are given in Section \ref{Prel}.
An important functional, which attains maximum at the global
Maxwellian function is introduced in Section \ref{Extr}.
The distance between solutions and the corresponding extremal problem
are studied in Section \ref{Dist}.

We use the notation $C_{0}^{1}$ to denote the class
of differentiable functions $f(\zeta)$, which tend to zero sufficiently rapidly
when $\zeta$ tends to infinity.
 \section{Preliminaries: main definitions and results}\label{Prel}
In this section we  present some  well-known  notions and results
 connected with the Boltzmann equation.
The distribution function $f(t,x,\zeta)$ is non-negative:
\begin{equation}\label{3} f(t,x,\zeta){\geq}0,
\end{equation}
and so the entropy
 \begin{equation}\label{4}
 S(t,f)=-\int_{\Omega}\int_{R^{n}}f(t,x,\zeta)\log
  f(t,x,\zeta)d{\zeta}dx
\end{equation}
is well-defined.
\begin{Dn}\label{CInv}
A function $\phi(\zeta)$ is called a collision invariant if it satisfies
 the relation
 \begin{equation}\label{5}
  \int_{\BR^{n}}\phi(\zeta)Q(f,f)(\zeta)d\zeta=0\quad  {\mathrm{for \,\, all}}
  \quad
  f{\in}C_{0}^{1}.
 \end{equation}
\end{Dn}
 It is well-known (see \cite{Vi1}) that there are the following collision invariants:
 \begin{equation}\label{6}
 \phi_{0}(\zeta)=1,\quad
 \phi_{i}(\zeta)=\zeta_{i} \quad(i=1,2,...,n),\quad
  \phi_{n+1}(\zeta)=|\zeta|^{2}.
  \end{equation}
The notions of density $\rho(t,x)$,  total density $\rho(t)$, mean velocity $u(t,x)$,
energy $E(t,x)$, and total  energy $E(t)$ are introduced via formulas:
 \begin{align}\label{7}&
 \rho(t,x)=\int f(t,x,\zeta)d\zeta , \quad \rho(t)=\int_{\Omega}\rho(t,x){dx},\\
 \label{8}&
 u(t,x)=\big(1/\rho(x,t)\big)\int \zeta f (t,x,\zeta){d}\zeta,
 \\ \label{9}&
 E(t,x)=\int \frac{|\zeta|^{2}}{2}f(t,x,\zeta)d\zeta , \quad
E(t)=\int_{\Omega}\int_{\BR^n} \frac{|\zeta|^{2}}{2}f(t,x,\zeta)d\zeta dx.
\end{align}
The function
\begin{equation}\label{9'}
f=\big(\rho / (2{\pi}T)^{n/2} \big) \exp\big(- |\zeta-u|^{2} /(2T)\big).
\end{equation}
is called the \emph{global Maxwellian} and is a function of the mass density $\rho>0$, bulk velocity $u=(u_{1},...,u_{n})$ and temperature T.
We assume that the domain $\Omega$ is bounded and so its volume
is bounded too:
\begin{equation}\label{11}
\Vol(\Omega)=V_{\Omega}<\infty.
\end{equation}
Therefore, the function
\begin{equation}\label{9''}
M(\zeta)=\big(\rho / \big(V_{\Omega}(2{\pi}T)^{n/2} \big)\big)
 \exp\big(- |\zeta-u|^{2} /(2T)\big)
\end{equation}
is a global Maxwellian with the constant total density $\rho$.
\begin{Pn}\label{Pn1} \cite{Vi1} The global Maxwellian function $M(\zeta)$ is the stationary solution of the Boltzmann equation \eqref{1}.
\end{Pn}
Boltzmann proved  in \cite{Bol} the  fundamental result below:
\begin{Tm} \label{Tm2} Let $f\in C_{0}^{1}$ be a non-negative solution of equation \eqref{1}. Then the following inequality holds:
\begin{equation}\label{10}
{dS}/{dt}\,\, {\geq} \,\, 0.
\end{equation} 
\end{Tm}
\section{Extremal problem}\label{Extr}
Similar to the cases considered in \cite{LAS3, LAS4},
 an important role is played by the functional
\begin{equation}\label{12}
 F(f)={\lambda}{E}+S,\quad \lambda=-1/T,\end{equation}
 where $S$ and ${E}$, respectively, are defined by formulas \eqref{4} and \eqref{9}.  The parameters $\lambda=-1/T$ and $\rho$ are fixed. Now,   we use the calculus
of variations (see \cite{Ha}) and find the function $f_{\max}$ which maximizes
the functional \eqref{12} on  the class of functions with the same $\rho(t)=\rho$ at the fixed moment $t$.
 The corresponding Euler's equation takes the form
\begin{equation}\label{14}
\frac{\delta}{\delta{f}}\Big[{\lambda}\frac{|\zeta|^{2}}{2}f-f\log f+
{\mu}f\Big]=0.\end{equation}\label{15}
Here $\frac{\delta}{\delta{f}}$ stands for the functional derivative.
Our extremal problem is conditional and
 $\mu$ is the Lagrange multiplier. Hence, we have
\begin{equation}\label{16}
{\lambda}\frac{|\zeta|^{2}}{2}-1- \log{f}+
{\mu}=0.
\end{equation}
From the last relation we obtain
\begin{equation}\label{17}
f=Ce^{-|\zeta|^{2}/(2T)}.
\end{equation}
Formulas \eqref{9''} and \eqref{17}
imply that
\begin{equation}\label{18}
f=M(\zeta)=\frac{\rho}{V_{\Omega}(2{\pi}T)^{n/2}}e^{-\frac{|\zeta|^{2}}{2T}}.
\end{equation}
We have the inequality
\begin{equation}\label{19}
\frac{\delta^{2}}{\delta{f}^{2}}F=-1/f<0.
\end{equation}
\begin{Cy}\label{Cy}
The global Maxwellian function $M(\zeta)$, which is
defined by formula \eqref{17}, gives the maximum of the functional  $F$
on the class of functions with the same value $\rho$
of the total density $\rho(t)$ at the fixed moment $t$.
\end{Cy}
In view of \eqref{4}, \eqref{9}, and \eqref{12} we see that
\begin{equation}\label{!}
\big(F(f)\big)(t)=-\int_{\Omega}\int_{\BR^{n}}
\Big(\frac{|\zeta|^{2}}{2T}+\log f(t,x,\zeta)\Big)f(t,x,\zeta)d{\zeta}dx.
\end{equation}
It follows from \eqref{7}, \eqref{18}, and \eqref{!} that
\begin{equation}\label{4.3}
F(M)=-{\rho}\log\Big(\frac{\rho}{V_{\Omega}(2{\pi}T)^{n/2}}\Big).
\end{equation}
Therefore, Corollary \ref{Cy}
can also be  proved without using the calculus of variation  (see \cite{TVi}).
Indeed, taking into account  
relations \eqref{18}, \eqref{!}, and \eqref{4.3} and the fact that
 the total densities of $M$ and $f$ are equal, we have
\begin{equation}\label{20}
F(M)-F(f)=\int_{\Omega}\int_{\BR^{n}}
M\Big(1-\frac{f}{M}+\frac{f}{M}\log\frac{f}{M}\Big)d{\zeta}dx.
\end{equation}
Using inequality $1-x+x\log{x}>0$ for $x>0,\,\, x{\ne}1$, we derive from
\eqref{20} that
\begin{equation}\label{2!}
F(M)-F(f)>0
\quad (f{\ne}M).
\end{equation}
\begin{Rk} \label{Rk3.1} Since the extremal problem is conditional, the connection between the energy and entropy can be interpreted in terms of game theory.
The  functional  \eqref{12} defines this game. The global Maxwellian function $M(\zeta)$ is the solution of it.  A game interpretation of quantum and classical mechanics problems is given in the papers \cite{LAS3, LAS4}.
\end{Rk}
\section{Distance}\label{Dist}
Let $f(t,x,\zeta)$ be a nonnegative solution of the Boltzmann equation \eqref{1}. We assume that $T$ and the value
$\rho=\rho(t)$ at some moment $t$  are fixed. According to \eqref{2!} we have
\begin{equation} \label{4.1}
F(M)-F(f){\geq}0,
\end{equation}
where the global Maxwellian function $M(\zeta)$ is defined in \eqref{18}. The equality in  \eqref{4.1} holds
if and only if $f(t,x,\zeta)=M(\zeta)$. Hence, we can introduce the following
definition of distance between the solution $f(t,x,\zeta)$ and the global Maxwellian function $M(\zeta)$:
\begin{equation}\label{4.2}
\dist\{M,f\}=F(M)-F(f).
\end{equation}
\begin{Rk}\label{Rk4.1}  In the  spatially homogeneous case,
if not only the total densities
$\rho_{M}$ and $\rho_{f}$ of $M$ and $f$  are equal,
but the energies $E_{M}$ and $E_{f}$ are equal too, 
then our definition \eqref{4.2} of distance
coincides with the Kullback-Leibler distance (see \cite{Vi2}).
However, our approach enables us to treat also the non-homogeneous case.
\end{Rk}
Next, we study the case $E_M\ne E_f$ and start with an example.
\begin{Ee} \label{Ee4.2} 
Let $T_1\ne T$ and consider the global   Maxwellian function 
\begin{equation}\label{4.4}
M_{1}(\zeta)=\frac{\rho}{V_{\Omega}(2{\pi}T_{1})^{n/2}}
\exp\Big(-\frac{|\zeta|^{2}}{2T_{1}}\Big).
\end{equation}
Direct calculation shows that
\begin{align}\label{3!}&
E_1=E_{M_1}=\rho nT_1/2\not=E, \\
\label{4.5}&
F(M_{1})=-{\rho}\Big(\log \big(\frac{\rho}{V_{\Omega}(2{\pi}T_{1})^{n/2}}\big)-n(1-T_{1}/T)/2\Big).
\end{align}
It follows from \eqref{4.3} and \eqref{4.5} that
\begin{equation} \label{4.6}
\dist\{M,M_{1}\}=-{\rho}n\big(\log(T_{1}/T)-T_{1}/T+1\big)/2.
\end{equation}
\end{Ee}
We introduce the class $C(\rho,E_{1},U)$ of non-negative functions
functions $f(t,x,\zeta)$ with the given total density $\rho$ (see (2.8)), total energy
\begin{equation}\label{4.7}
\int_{\Omega}\int_{\BR^n}\frac{|\zeta|^{2}}{2}f(t,x,\zeta)d{\zeta}dx=E_1,
\end{equation}
and total moments $U=\big(U_{1},U_{2},...,U_{n}\big)$,
where
\begin{equation}\label{4.8}
U_{k}=\int_{\Omega}\int_{\BR^n}{\zeta_{k}}f(t,x,\zeta)d{\zeta}dx.
\end{equation}
Recall that  the global Maxwellian function $M$ is defined by \eqref{18}.

\textbf{Extremal problem.} \emph{Find  a function $f$, which  minimizes 
the functional} $\dist\{M,f\}$ \emph{on the class} $C(\rho,E_{1},U)$.

The corresponding Euler's equation takes the form
\begin{equation}\label{4.9}
\frac{\delta}{\delta{f}}\Big[({\lambda}+\nu)\frac{|\zeta|^{2}}{2}f-f\log f+
{\mu}f+f\sum_{k}\gamma_{k}\zeta_{k}\Big]=0.\end{equation}  
Recall that our extremal problem is conditional, and
 $\mu,\, \nu,\, \gamma_{k}$ are  the Lagrange multipliers. Hence, we have
\begin{equation}\label{4.10}
({\lambda}+\nu)\frac{|\zeta |^{2}}{2}-\log f-1
+{\mu}+\sum_{k}\gamma_{k}\zeta_{k}=0.
\end{equation}
From the last relation we obtain
\begin{equation}\label{4.11}
f=C\exp\Big((\lambda+\nu)\frac{|\zeta|^{2}}{2}+\sum_{k}\gamma_{k}\zeta_{k}\Big).
\end{equation}
According to \eqref{7} we have $\lambda +\nu<0$. Now, we rewrite \eqref{4.11} as
\begin{equation}\label{4!}
f=C_1\Big(-\frac{2\pi}{\lambda + \nu} \Big)^{-n/2}
\exp\Big(\frac{\lambda+\nu}{2}\sum_{k}\Big(\zeta_k+\frac{\gamma_{k}}
{\lambda+\nu}\Big)^2\Big),
\end{equation}
where
\begin{equation}\label{4.13}
C_{1}=C\frac{\pi^{n/2}}{(-(\lambda+\nu)/2)^{n/2}}\exp\Big(-\frac{\sum_{k}\gamma_{k}^{2}}{2(\lambda+\nu)}\Big).
\end{equation}
To calculate the parameters $\mu,\nu,\gamma_{k}$ we use again the well-known formulas
\begin{equation}\label{4.12}
\int_{-\infty}^{\infty}e^{-a\xi ^{2}}d\xi=\sqrt{\pi/a},\quad
\int_{-\infty}^{\infty}\xi^{2}e^{-a\xi^{2}}d\xi=\frac{1}{2a}\sqrt{\pi/a},\quad a>0.
\end{equation}
Formulas \eqref{7}, \eqref{4.7}, \eqref{4.8}, \eqref{4!}, and \eqref{4.12}
imply that
\begin{equation}\label{4.14}
C_{1}=\rho/V_{\Omega},\quad \gamma_{k}/(\lambda+\nu)=-U_{k}/\rho,
\quad -(\lambda+\nu)=T_{1}^{-1},
\end{equation}
where 
\begin{equation}\label{4.15}
T_1=\frac{2}{n\rho}E_1-\frac{1}{n\rho^2}\sum_k U_k^2.
\end{equation}
Because of \eqref{4!} and \eqref{4.14} we see that $f$ is just another
global Maxwellian function
\begin{equation}\label{4.17}
f=M_{1}(\zeta)=\frac{\rho}{V_{\Omega}(2{\pi}T_{1})^{n/2}}
\exp\Big(-\frac{|\zeta-U/{\rho}|^{2}}{2T_{1}}\Big).
\end{equation}
Moreover, the inequality
\begin{equation}\label{4.16}
\frac{\delta^{2}}{\delta{f}^{2}}[\dist\{M,f\}]=1/f
\end{equation}
holds, that is, the functional $\dist\{M,f\}$ attains its minimum on the function $f=M_1$, which satisfies conditions $\rho(t)=\rho$, \eqref{4.7}, and \eqref{4.8}. The following assertion is true.
\begin{Pn}
\label{Pn4.1} Let $M$ and $M_{1}$, respectively, be defined by \eqref{18} and \eqref{4.17}. If the function $f$ satisfies conditions $\rho(t)=\rho$,
\eqref{4.7}, \eqref{4.8}, and $f{\ne}M_{1}$, then
\begin{equation}\nn
\dist\{M,f\}>-\frac{n{\rho}}{2}\big(\log(T_{1}/T)-T_{1}/T+1\big)+\frac{|U|^2}{2\rho T_1}.
\end{equation}
\end{Pn}


\begin{thebibliography}{1111}
\bibitem{Bol}
Boltzmann L., \emph{Lectures on Gas Theory},
Courier Dover Publications, 1995. 
\bibitem{Hab}
Haba Z.,	\emph{Non-linear relativistic diffusions}
Physica A: Statistical Mechanics and its Applications, 
doi:10.1016/j.physa.2011.03.025
\bibitem{Ha}
Hahn W., \emph{Theory and Application of Liapunov's Direct Method}, Englewood Cliffs, NJ: Prentice-Hall, 1963.
\bibitem{KL}
Kullback S., Leibler R.A.,
\emph{On information and sufficiency},
Ann. Math. Stat. 22, 79-86, 1951.
\bibitem{LAS1}
Sakhnovich L.A., \emph{Comparing Quantum and Classical Approaches in
Statistical Physics}, Theor. Math. Phys. 123:3, 846-850, 2000.
\bibitem{LAS2}
Sakhnovich L.A., \emph{Comparison of Thermodynamic Characteristics of
a Potential Well under Quantum and Classical Approaches},
Funct. Anal. Appl. 36:3, 205-211, 2002.
\bibitem{LAS3}
Sakhnovich L.A., \emph{Comparison of Thermodynamics Characteristics in Quantum
and Classical Approaches and Game Theory}, arXiv:10104717, v.2, 
Physica A to appear.
\bibitem{LAS4}
Sakhnovich L.A., \emph{Laws of thermodynamics and game theory},
arXiv:1105.4633.
\bibitem{SH}
Sobczyk K., Holobut P.,
\emph{Information-theoretic approach to dynamics of stochastic systems},
Probabilistic Engineering Mechanics, doi:10.1016/j.probengmech.2011.05.007
\bibitem{TVi}
Toscani G., Villani C.,
\emph{Sharp entropy dissipation bounds and explicit rate of trend to equilibrium for the spatially homogeneous Boltzmann equation},
Comm. Math. Phys. 203:3, 667Ð706, 1999.
\bibitem{Vi1}
Villani C., \emph{A review of mathematical topics in collisional kinetic theory},
in: Handbook of mathematical fluid dynamics, Vol. I, 71Ð305, Amsterdam:
North-Holland, 2002. 
\bibitem{Vi2}
Villani C., \emph{Entropy production and convergence to equilibrium
for the Boltzmann equation},  in:
Zambrini J.-C. (ed.), XIVth international congress on mathematical physics. Selected papers,  130-144, Hackensack, NJ: World Scientific,  2005. 
\end{thebibliography}
\end{document}